\documentclass[pra,a4paper,showpacs,superscriptaddress]{revtex4-1}
\usepackage{amsmath}
\usepackage{amsfonts}
\usepackage{graphicx}
\usepackage{longtable}
\usepackage{multirow}
\newcommand{\be}{\begin{equation}}
\newcommand{\ee}{\end{equation}}
\newcommand{\bee}{\begin{equation*}}
\newcommand{\eee}{\end{equation*}}

\newcommand{\ba}[1]{\left(\begin{array}{#1}}
\newcommand{\ea}{\end{array}\right)}
\begin{document}

\title{ONE PARAMETER FAMILY OF $N$-QUDIT WERNER-POPESCU STATES: BIPARTITE SEPARABILITY USING CONDITIONAL  
QUANTUM RELATIVE TSALLIS ENTROPY}

\author{Anantha S Nayak}
\affiliation{Department of Physics, Kuvempu University, 
Shankaraghatta, Shimoga-577 451, India} 
\author{Sudha }
\affiliation{Department of Physics, Kuvempu University, 
Shankaraghatta, Shimoga-577 451, India}
\affiliation{Inspire Institute Inc., Alexandria, Virginia, 22303, USA.}
\author{A. R. Usha Devi}
\affiliation{ Department of Physics, Bangalore University, Bangalore 560 056, India}
\affiliation{Inspire Institute Inc., Alexandria, Virginia, 22303, USA.}

\author{A. K. Rajagopa1}\affiliation{Inspire Institute Inc., Alexandria, Virginia, 22303, USA.}
\affiliation{Harish-Chandra Research Institute, Chhatnag Road, Jhunsi, Allahabad 211 019, India}
\affiliation{Institute of Mathematical Sciences, C.I.T. Campus, Taramani, Chennai, 600113, India} 

\date{\today}

\begin{abstract}
The conditional version of sandwiched Tsallis relative entropy (CSTRE) is employed to study the bipartite separability of one parameter family of $N$-qudit Werner- Popescu states in their $1 : N-1$ partition. For all $N$, the strongest limitation on bipartite separability is realized in the limit  $q\rightarrow \infty$ and is found to match exactly with the  separability range obtained using an algebraic method which is both necessary and sufficient. The theoretical superiority of using CSTRE criterion to find the bipartite separability range over the one using Abe- Rajagopal (AR) $q$-conditional entropy is illustrated by comparing the convergence of the parameter $x$ with respect to $q$, in the implicit plots of AR $q$-conditional entropy and CSTRE.   
\end{abstract}

\keywords{Entropic separability criterion; $q$-conditional entropies; Non-commuting version of relative entropy}
\maketitle
\section{Introduction}
Entropic characterization of separability~$[1--21]$ in mixed composite states has witnessed a considerable interest in recent years~$[17--21]$. The identification of a non-commuting generalization of Abe-Rajagopal (AR) $q$-conditional Tsallis entropy\cite{e3}  in the form of conditional version of sandwiched Tsallis relative entropy (CSTRE)~\cite{asn} and its usefulness in identifying a separability range stricter than the separability criterion using AR q-conditional entropy ({\emph{the so-called AR-criterion}})\cite{e3}, has given more impetus to this study\cite{asn,asn2,asn3}. It has been established that negative values of CSTRE imply entanglement in  chosen bipartitions of any composite state~\cite{asn2}.  In noisy one-parameter families of symmetric\cite{asn2} and non-symmetric\cite{asn3} multiqubit states the separability range obtained through CSTRE, in addition to being stricter than that through AR-criterion, is shown to match with the separability range obtained through Peres' Partial Transpose 
(PPT) criterion\cite{peres,horodecci}. While AR-criterion\cite{e3,e4,e5,e8,e9} relies upon the local versus global disorder thus exhibiting its spectral nature, the CSTRE criterion is illustrated to have non-spectral features\cite{asn}.     

Quite similar to the definition of sandwiched  R\'{e}nyi relative 
entropy\cite{e11,e12,e13}, the sandwiched form of the Tsallis relative entropy is identified to be\cite{asn}
\be
\label{cstre1}
\widetilde{D}^{T}_{q}(\rho || \sigma ) = \frac{\mbox{Tr}\left\{
\left(\sigma ^{\frac{1-q}{2q}} \rho \  \sigma ^{\frac{1-q}{2q}}\right)\right\}^{q}-1}{q-1}
\ee
Eq. (\ref{cstre1}) reduces to traditional relative Tsallis entropy $ D^{T}_{q}(\rho || \sigma ) $
\be
\label{ttre}
D^{T}_{q}(\rho || \sigma ) = \frac{\mbox{Tr}\left[\rho^{q} \ \ \sigma^{1-q}\right]-1} {q-1} 
\ee
when $ \rho $ and $ \sigma $  commute with each other.  

The conditional forms of $ \widetilde{D}^{T}_{q}(\rho || \sigma ) $ are defined  as\cite{asn} 
\be
\label{tcsre}
\widetilde{D}^{T}_{q} \left( \rho_{AB} || \rho_{B} \right) = \frac{\widetilde{Q}_{q} \left( 
\rho_{AB} || \rho_{B} \right) - 1 } {1-q}
\ee 
and
\be
\label{tcsreA}
\widetilde{D}^{T}_{q} \left( \rho_{AB} || \rho_{A} \right) = \frac{\widetilde{Q}_{q} \left( 
\rho_{AB} || \rho_{A} \right) - 1 } {1-q}
\ee
with $\widetilde{Q}_{q} \left(\rho_{AB} || \rho_{B} \right)$, $\widetilde{Q}_{q} \left( 
\rho_{AB} || \rho_{A} \right)$ being respectively given by 
\begin{eqnarray}
\label{qsf}
\widetilde{Q}_{q} \left( \rho_{AB} || \rho_{B} \right)& =& \mbox{Tr} \left\{\left[\left(I_{A} \otimes 
\rho_{B}\right)^{\frac{1-q}{2q}} \rho_{AB} \ \left(I_{A} \otimes \rho_{B}\right)^{\frac{1-q}{2q}}\right] ^{q} \right\}  \\
& & \nonumber \\
\label{qsfA}
\widetilde{Q}_{q} \left( \rho_{AB} || \rho_{A} \right)& =& \mbox{Tr} \left\{\left[\left(\rho_{A} \otimes 
I_{B}\right)^{\frac{1-q}{2q}} \rho_{AB} \ \left(\rho_{A} \otimes I_{B}\right)^{\frac{1-q}{2q}}  \right] ^{q} \right\}
\end{eqnarray}
In Ref. \cite{asn2}, it has been proved that negative values of 
$\widetilde{D}^{T}_{q} \left( \rho_{AB} || \rho_{B} \right)$, $\widetilde{D}^{T}_{q} \left( \rho_{AB} || \rho_{A} \right)$  indicate entanglement in the bipartite state $\rho_{AB}$.  
When the subsystems $ \rho_{B}$ or $\rho_A$ are maximally mixed,  Eqs. (\ref{tcsre}), (\ref{tcsreA})) reduce to Abe-Rajagopal (AR) q-conditional Tsallis entropies\cite{e3} $S^{T}_{q}(A|B)$, $S^{T}_{q}(B|A)$ respectively: 
\begin{eqnarray}
\label{arre}
S^{T}_{q}(A|B)&=& \frac{1}{q-1} \left[1-  \frac{\mbox{Tr} \rho^{q}_{AB}}{\mbox{Tr} \rho^{q}_{B}}\right], \\ 
& & \nonumber \\
\label{arreA}
S^{T}_{q}(B|A)&=& \frac{1}{q-1} \left[1-  \frac{\mbox{Tr} \rho^{q}_{AB}}{\mbox{Tr} \rho^{q}_{A}}\right].
\end{eqnarray} 
Quite like the AR $q$-conditional entropies $S^{T}_{q}(A|B)$, $S^{T}_{q}(B|A)$, both the conditional versions of sandwiched Tsallis relative entropy $\widetilde{D}^{T}_{q} \left( \rho_{AB} || \rho_{B} \right)$, $\widetilde{D}^{T}_{q} \left( \rho_{AB} || \rho_{A} \right)$ reduce to the respective von-Neumann entropies 
$S(A\vert B)$, $S(B\vert A)$ in the limit $q\longrightarrow 1$.

Both AR- and CSTRE- criteria have been employed in Refs. ~\cite{asn,asn2} to find the $1:N-1$ separability range of the  noisy one parameter families of symmetric $N$- qubit states involving either W or GHZ states. In Ref.~\cite{asn3}, the $1:N-1$ separability ranges in two different non-symmetric one-parameter families of $N$-qubit states are obtained using 
AR-, CSTRE criteria  and a comparative analysis of these separability ranges is carried out. 

The investigation of  separability range in one parameter families of mixed states through 
AR- and CSTRE criteria has revealed that whenever the marginal is not maximally mixed and hence does not commute with the global density matrix, the CSTRE criterion yields stricter separability range than its commuting version, the AR-criterion\cite{hhh,hhh2,asn,asn2,asn3}. If the marginal is maximally mixed thus commuting with its density matrix, both AR-, CSTRE-criteria are found to yield identical separability ranges\cite{asn,asn2,asn3}. The supremacy of CSTRE criterion over AR-criterion, in the cases where non-maximal marginals occur, is illustrated for symmetric\cite{asn,asn2} and non-symmetric one-parameter families of multiqubit states\cite{asn3}.   In this work, we wish to examine whether CSTRE criterion remains superior to 
AR-criterion even for composite states containing {\emph {qudits}}.  For this purpose, we have considered $N$-qudit Werner-Popescu states\cite{e5}, 
a special one parameter family of states and examine its $1:N-1$ separability range using CSTRE criterion.  
Both AR-, CSTRE-criteria are seen to result in the necessary and sufficient condition for separability in the $1:N-1$ partition of these states. Further we compare the convergence of the parameter $x$, obtained through CSTRE criterion with that obtained through AR criterion, with respect to $q$. It has been observed that the parameter $x$ converges rapidly in the case of AR criterion, in comparison with that in the case of CSTRE criterion, even for finite values of $q$  thus implying the better stochasticity of CSTRE criterion over the AR-criterion. 

This article is organized in four sections including the introductory section (Section~1) in which we recall the non-additive entropic separability criteria  such as AR-, CSTRE-criteria and discuss the motivation behind this work.  Section~2 introduces the $N$-qudit Werner-Popescu state as a generalization of noisy one-parmeter family of $N$-qubit GHZ states to its qudit counterpart.  Section~3  examines the $1:N-1$ separability range of one parameter family of $N$-qudit Werner-Popescu states using different separability criteria. A comparison of the results obtained through AR-, CSTRE criteria are compared and the superiority of CSTRE criterion is illustrated through the implicit plots of $x$ versus $q$ in both AR-, CSTRE methods (Section~3). Finally Section 4 provides a summary of the results. 

\section{$N$-qudit Werner-Popescu states} 
\label{ch5:wpN}
The Werner-Popescu state with $N$-qudits\cite{e5} is defined as
\begin{eqnarray}
\nonumber \rho^{d}_{N} (x) & = & \rho\left( A_{1},\,A_{2},\, \ldots A_{N}\right) \nonumber \\
& & \nonumber \\
&=&\frac{1-x}{d^{N}}\left[ I_{d} (A_{1}) \otimes I_{d} (A_{2}) \otimes \ldots I_{d} (A_{N})\right] + \ x \left|\Phi^{N}_{d}\right\rangle \ \left\langle \Phi^{N}_{d}\right|  
\end{eqnarray}
Here  $0\leq x\leq 1$ and $ I_{d} (A_i)$, $i=1,\,2,\ldots, N$ are  $ d \times d $ unit matrices belonging to the subsystem space of each qudit $A_i$, $i=1,\,2,\ldots, N$. The pure state 
$\left|\Phi^{N}_{d}\right\rangle $ is given by 
\be
\label{dghz}
\left|\Phi^{N}_{d}\right\rangle = \frac{1}{\sqrt{d}} \ \sum ^{d-1}_{k=0}  \ \left|k\right\rangle _{A_{1}} \otimes \left|k\right\rangle _{A_{2}} \otimes \ldots \otimes \left|k\right\rangle _{A_{N}}.
\ee
and it is an analogue of GHZ state to $d$-level systems.  Notice that when $d=2$, i.e., for qubits, 
$k=0$, $1$ and Eq. (\ref{dghz}) reduces to the $N$-qubit GHZ state
\[
\vert {\rm GHZ}_N \rangle=\frac{1}{\sqrt{2}} \left( \vert 0_10_2\cdots 0_N \rangle + 
\vert 1_1 1_2\cdots 1_N \rangle   \right)  
\]  
The eigenvalues of $\rho^d_N(x)$ are given by 
\begin{eqnarray}
\lambda_1&=&  \frac{1-x}{d^{N}} \ \ \ \ \ \ \ \ \ \ \ \   [(d^{N}-1) \mbox{fold degenerate}], \nonumber \\
& & \nonumber \\
\lambda_2&=&\frac{1+(d^{N}-1)x}{d^{N}} \ \ \ \ \ \ \ \  \mbox{non-degenerate} 
\end{eqnarray}
The focus here is to  find the $1:N-1$ separability range of $\rho^d_N(x)$ using CSTRE criterion.

\section{Bipartite separability of $\rho^{d}_{N}(x)$ in its $1:N-1$ partition} 

Denoting the first qubit as subsystem $A$ and the remaining $N-1$ qubits as subsystem $B$, the density 
matrix of the $N-1$ qubit marginal is given by
\begin{eqnarray}
\nonumber \rho_{B}&=&=\mbox{Tr}_{A_{1}}\,\rho\left( A_{1},\,A_{2},\ldots,A_{N}\right)=\mbox{Tr}_{A_{1}}\,\rho^d_N(x) 
\end{eqnarray}
It can be seen that the eigenvalues $\eta_i$ of  the $N-1$ qubit marginal $\rho_B$ of $\rho_N^d(x)$, obtained by reducing over the first qubit, 
are  given by
\begin{eqnarray}
\eta_1 &= & \frac{1-x}{d^{N-1}} \ \ \ \ \ \ \ \ \ \ \ \    [(d^{N-1}-d)- \mbox{fold degenerate}], \nonumber \\
& & \nonumber \\
\eta_2 &=& \frac{1+(d^{N-2}-1)x}{d^{N-1}} \ \ \ \ \ \ \ \ \   [d- \mbox{fold degenerate}] 
\end{eqnarray}
Also, the subsystem $\rho_A$, the single qudit marginal  of $\rho^d_N(x)$, corresponds to the maximally mixed state $I_d/d$, $I_d$ being $d\times d$ unit matrix. 

In order to find the separability range of the state $\rho_N^d$ in its $ 1: N-1 $ partition using CSTRE criterion, one needs to evaluate the eigenvalues $\gamma_{i}$ of the sandwiched matrix  
\be
\Gamma=\left(I_{A} \otimes \rho_B\right)^{\frac{1-q}{2q}} \rho^d_{N}(x) \ \left(I_{A} \otimes 
\rho_B\right)^{\frac{1-q}{2q}}
\ee 
so that  
\be
\widetilde{D}^{T}_{q} \left( \rho_{N}^d (x)|| \rho_B \right) = \frac{\sum_{i} \gamma^{q} _{i} - 1 } {1-q}
\ee
can be evaluated. Thus, in the evaluation of $\widetilde{D}^{T}_{q} \left( \rho_{N}^d(x) || \rho_B \right)$, the non-negative eigenvalues $\gamma_{i}$ play a crucial role. In order to  obtain the form of the eigenvalues $\gamma_i$ for arbitrary $N$, an analysis of their form for different 
$N (N= 2,\,3,\,4,\,5)$ and $d \ (d=3,\,4,\,5,\,6)$ is carried out to arrive at a  generalization for any $N$, $d$. Table \ref{ch5table} provides the explicitly evaluated non-zero eigenvalues of the sandwiched matrix $\Gamma$ for different values of $N$ and $d$.
\begin{table}[ht]
\caption{The non-zero eigenvalues $\lambda_{i}$ of the sandwiched matrix 
$\left(I_{A} \otimes \rho_{B}\right)^{\frac{1-q}{2q}} \rho^{d}_{N}(x) \ \left(I_{A} \otimes \rho_{B}\right)^{\frac{1-q}{2q}} $. }
\label{ch5table}
\begin{tabular}{|c|c|c|c|c|}
\hline
Number  & Number & $\gamma_{1} $  & $\gamma_{2}$ & $\lambda_{3}$ \\ 
of & of  & $\left(d^{N}-d^{2}\right)$ fold & $\left(d^{2}-1\right)$ fold & \\
 levels ($d$) & parties ($N$) & degenerate & degenerate & \\
\hline\hline 
\hline\hline 
\multirow{4} {*}  {3}  &  2  & - & $\left(\frac{1-x}{9}\right) \left(\frac{1}{3}\right)^{\frac{1-q}{q}}$ & $\left(\frac{1+8x}{9}\right) \left(\frac{1}{3}\right)^{\frac{1-q}{q}}$ \\ \cline{2-5} 
  &  3  & $\left(\frac{1-x}{27}\right) \left(\frac{1-x}{9}\right)^{\frac{1-q}{q}}$ & $\left(\frac{1-x}{27}\right) \left(\frac{1+2x}{9}\right)^{\frac{1-q}{q}}$ & $\left(\frac{1+26x}{27}\right) \left(\frac{1+2x}{9}\right)^{\frac{1-q}{q}}$ \\ \cline{2-5} 
 & 4 &  $\left(\frac{1-x}{81}\right) \left(\frac{1-x}{27}\right)^{\frac{1-q}{q}}$  & $\left(\frac{1-x}{81}\right) \left(\frac{1+8x}{27}\right)^{\frac{1-q}{q}}$ & $\left(\frac{1+80x}{81}\right) \left(\frac{1+8x}{27}\right)^{\frac{1-q}{q}} $  \\ \cline{2-5} 
 &  5  & $\left(\frac{1-x}{243}\right) \left(\frac{1-x}{81}\right)^{\frac{1-q}{q}}$ & $\left(\frac{1-x}{243}\right) \left(\frac{1+26x}{81}\right)^{\frac{1-q}{q}}$ & $\left(\frac{1+242x}{243}\right) \left(\frac{1+26x}{81}\right)^{\frac{1-q}{q}}$ \\ \cline{1-5} 
\multirow{4} {*} {4}  &  2  & - &  $\left(\frac{1-x}{16}\right) \left(\frac{1}{4}\right)^{\frac{1-q}{q}}$ & $\left(\frac{1+15x}{16}\right) \left(\frac{1}{4}\right)^{\frac{1-q}{q}} $ \\ \cline{2-5} 
  &  3  & $\left(\frac{1-x}{64}\right) \left(\frac{1-x}{16}\right)^{\frac{1-q}{q}} $ & $\left(\frac{1-x}{64}\right) \left(\frac{1+3x}{16}\right)^{\frac{1-q}{q}}$ & $\left(\frac{1+63x}{64}\right) \left(\frac{1+3x}{16}\right)^{\frac{1-q}{q}}$ \\ \cline{2-5} 
 &  4  & $\left(\frac{1-x}{256}\right) \left(\frac{1-x}{64}\right)^{\frac{1-q}{q}}$ & $\left(\frac{1-x}{256}\right) \left(\frac{1+15x}{64}\right)^{\frac{1-q}{q}}$ & $\left(\frac{1+255x}{256}\right) \left(\frac{1+15x}{64}\right)^{\frac{1-q}{q}}$ \\ \cline{2-5} 
 &  5  & $\left(\frac{1-x}{1024}\right) \left(\frac{1-x}{256}\right)^{\frac{1-q}{q}}$ & $\left(\frac{1-x}{1024}\right) \left(\frac{1+63x}{256}\right)^{\frac{1-q}{q}}$ & $\left(\frac{1+1023x}{1024}\right) \left(\frac{1+63x}{256}\right)^{\frac{1-q}{q}}$ \\ \cline{1-5} 
\multirow{4} {*} {5}  &  2  & - &  $\left(\frac{1-x}{25}\right) \left(\frac{1}{5}\right)^{\frac{1-q}{q}}$ & $\left(\frac{1+24x}{25}\right) \left(\frac{1}{5}\right)^{\frac{1-q}{q}}$ \\ \cline{2-5} 
  &  3  & $\left(\frac{1-x}{125}\right) \left(\frac{1-x}{25}\right)^{\frac{1-q}{q}}$ & $\left(\frac{1-x}{125}\right) \left(\frac{1+4x}{25}\right)^{\frac{1-q}{q}}$ & $\left(\frac{1+124x}{125}\right) \left(\frac{1+4x}{25}\right)^{\frac{1-q}{q}}$ \\ \cline{2-5} 
  &  4 & $\left(\frac{1-x}{625}\right) \left(\frac{1-x}{125}\right)^{\frac{1-q}{q}}$ & $\left(\frac{1-x}{625}\right) \left(\frac{1+24x}{125}\right)^{\frac{1-q}{q}}$ & $\left(\frac{1+624x}{625}\right) \left(\frac{1+24x}{125}\right)^{\frac{1-q}{q}}$ \\ \cline{2-5} 
  &  5  & $\left(\frac{1-x}{3125}\right) \left(\frac{1-x}{625}\right)^{\frac{1-q}{q}}$ & $\left(\frac{1-x}{3125}\right) \left(\frac{1+124x}{625}\right)^{\frac{1-q}{q}}$ & $\left(\frac{1+3124x}{3125}\right) \left(\frac{1+124x}{625}\right)^{\frac{1-q}{q}}$ \\ \cline{1-5} 
\multirow{4} {*} {6}  &  2  & - &  $\left(\frac{1-x}{36}\right) \left(\frac{1}{6}\right)^{\frac{1-q}{q}}$ & $\left(\frac{1+35x}{36}\right) \left(\frac{1}{6}\right)^{\frac{1-q}{q}}$ \\ \cline{2-5} 
  &  3  & $\left(\frac{1-x}{216}\right) \left(\frac{1-x}{36}\right)^{\frac{1-q}{q}}$ & $\left(\frac{1-x}{216}\right) \left(\frac{1+5x}{36}\right)^{\frac{1-q}{q}}$ & $\left(\frac{1+215x}{216}\right) \left(\frac{1+5x}{36}\right)^{\frac{1-q}{q}}$ \\ \cline{2-5} 
 &  4  & $\left(\frac{1-x}{1296}\right) \left(\frac{1-x}{216}\right)^{\frac{1-q}{q}}$ & $\left(\frac{1-x}{1296}\right) \left(\frac{1+35x}{216}\right)^{\frac{1-q}{q}}$ & $\left(\frac{1+1295x}{1296}\right) \left(\frac{1+35x}{216}\right)^{\frac{1-q}{q}}$ \\ \cline{2-5} 
 &  5  & $\left(\frac{1-x}{7776}\right) \left(\frac{1-x}{1296}\right)^{\frac{1-q}{q}}$ & $\left(\frac{1-x}{7776}\right) \left(\frac{1+215x}{1296}\right)^{\frac{1-q}{q}}$ & $\left(\frac{1+7775x}{7776}\right) \left(\frac{1+215x}{1296}\right)^{\frac{1-q}{q}}$ \\ \cline{1-5} 
\end{tabular}
\end{table} 
It can be readily seen from Table \ref{ch5table} that, there are only three distinct non-zero eigenvalues for the sandwiched matrix 
$\Gamma$. A careful observation of the eigenvalues $\gamma_{i}, i=1,\,2,\,3 $ in Table \ref{ch5table} leads  towards the  generalization of the eigenvalues of sandwiched matrix $\Gamma$  for $N\geq 2$.   The generalized eigenvalues $\gamma_i$ of the sandwiched matrix $\Gamma$ for any $N\geq 2$ are given in the following:
\begin{eqnarray}
\nonumber \gamma_{1} &=& \left(\frac{1-x}{d^{N}}\right) \left(\frac{1-x}{d^{N-1}}\right)^\frac{1-q}{q}, \ \ 
\ \ \ \left(d^{N}-d^{2}\right)- \mbox{fold degenerate} \\
& & \nonumber \\
\nonumber \gamma_{2} &=&  \left(\frac{1-x}{d^{N}}\right) \left(\frac{1+\left(d^{N-2}-1\right)x}{d^{N-1}}\right)^\frac{1-q}{q}, 
\ \  \ \ \ \ \left(d^{2}-1\right)-\mbox{ fold degenerate} \\
& & \nonumber \\
\gamma_{3} &=&  \left(\frac{1+\left(d^{N}-1\right)x}{d^{N}}\right) \left(\frac{1+\left(d^{N-2}-1\right)x}{d^{N-1}}\right)^\frac{1-q}{q}, \ \ \mbox{non-degenerate}.
\end{eqnarray}
The $1: N-1$ separability range of $\rho^{d}_{N}(x)$,  for each combination of $ N=2,\,3,\,4,\,5$ and $ d=3,\,4,\,5,\,6 $ obtained using CSTRE approach allows us to generalize this range to any $N$ and $d$.  Table \ref{ch5wptable} gives the values of $x$ below which the state $\rho^{d}_{N}(x)$, ($ N=2,\,3,\,4,\,5$ and $ d=3,\,4,\,5,\,6 $) is separable. 
\begin{table}[ht]
\caption{ The comparison of the $1:N-1$ separability range of the state $\rho^{d}_{N} (x) $, for various compositions of  $d$ and $N$ obtained through CSTRE criterion. }
\label{ch5wptable}
\begin{tabular}{|c|c|c|}
\hline
Number  & Number &  CSTRE  \\ 
of & of  &  separability   \\
 levels ($d$) & parties ($N$)  &  range\\
\hline\hline 
\multirow{4} {*}  {3} & 2 & $\left(0, \ 0.25 \right)$  \\ \cline{2-3} 
  &  3  & $\left(0, \ 0.1\right)$ \\ \cline{2-3} 
 & 4 &  $\left(0, \ 0.0357 \right)$  \\ \cline{2-3} 
 & 5 & $\left(0, \ 0.0121\right)$  \\ \cline{1-3} 
\multirow{4} {*} {4}  &  2  & $\left(0, \  0.2\right)$  \\ \cline{2-3} 
  &  3  & $\left(0, \ 0.0588 \right)$  \\ \cline{2-3} 
 &  4  & $\left(0, \ 0.0153\right)$  \\ \cline{2-3} 
 &  5  & $\left(0, \ 0.0039 \right)$  \\ \cline{1-3} 
\multirow{4} {*} {5}  &  2  & $\left(0, \ 0.1666\right)$  \\ \cline{2-3} 
  &  3  & $\left(0, \ 0.0384 \right)$  \\ \cline{2-3} 
  &  4 & $\left(0, \ 0.0079\right)$ \\ \cline{2-3} 
  &  5  & $\left(0, \ 0.0016\right)$  \\ \cline{1-3} 
\multirow{4} {*} {6}  &  2  & $\left(0, \ 0.1428\right)$ \\ \cline{2-3} 
  &  3  & $\left(0, \ 0.0270\right)$  \\ \cline{2-3} 
 &  4  & $\left(0, \ 0.0046\right)$  \\ \cline{2-3} 
 &  5  & $\left(0, \ 0.0007\right)$\\ \cline{1-3} 
\end{tabular}
\end{table}
Using Table \ref{ch5wptable}, the following  
$1: N-1$ separability range is conjectured for the one parameter family of 
$N$-qudit Werner-Popescu-states.
\be
\label{ser}
0\leq x\leq \frac{1}{1+d^{N-1}}
\ee
One can note that the $1:N-1$ separability range given in Eq.(\ref{ser}) is the  
same as that obtained in Ref.~\cite{e5}, using the AR-criterion. In fact, the existence of maximally mixed single qubit density matrix is the reason 
behind the equivalence of separability ranges in CSTRE and AR-criteria. Such a situation occurs in the case of 
symmetric one parameter family of noisy GHZ states\cite{asn2}, psuedopure family containing GHZ states 
and Werner-like family of states containing GHZ states\cite{asn3}, while determining their $1:N-1$ separability range.  In all these states, the single qubit density matrix turns out to be $I_{2}/2$ thus commuting with the corresponding density matrix implying that the in general non-commutative CSTRE approach yields the results equivalent to commutative AR-approach\cite{asn2}. It is important to notice here that, using algebraic methods\cite{pitt1,pitt2} it has been shown that Eq.(\ref{ser}) is actually the necessary and sufficient condition for separability. 
 
\begin{figure}[ht]
\centerline{\includegraphics* [width=4in,keepaspectratio]{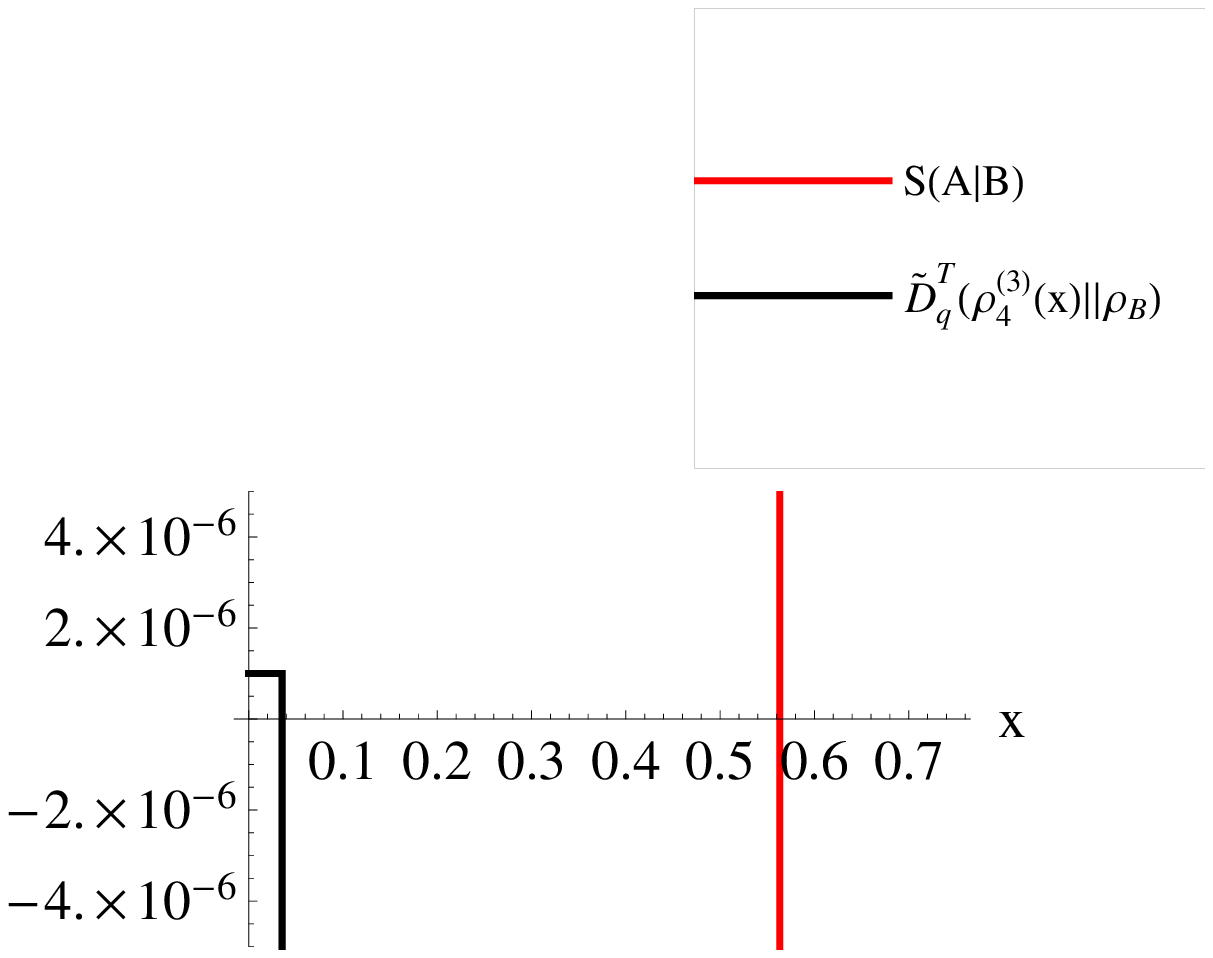}} 
\caption{The variation of conditional form of sandwiched Tsallis relative entropy $\tilde{D}^{T}_q(\rho^{(3)}_4(x)\vert\vert \rho_{B})$ in the $1:3$ partition of $4$-qutrit Werner-Popescu  states $\rho^{(3)}_4(x)$
($N$=4, $d=3$), with respect to $x$, in the limit $q\rightarrow \infty$.}
\label{abe34cstre}
\end{figure}

Fig.\ref{abe34cstre} gives an illustration of the monotonic decrease of $\tilde{D}^{T}_q(\rho^{(3)}_4(x)\vert\vert \rho_{B})$  with increasing $x$ in the limit
 $q\rightarrow \infty$.  

It can be seen that $\tilde{D}^{T}_q(\rho^{(3)}_4(x)\vert\vert \rho_{B})$ is negative for $ x > 0.5633 $ when $ q=1$ implying that $(0,\,0.5633)$ is the separability range through Von-Neumann conditional entropy, whereas it is negative for $ x > 0.0357 $ in the limit $ q \rightarrow \infty $ leading to $(0,\,0.357)$ as the separability range through CSTRE criterion.

\begin{figure}[ht]
\centerline{\includegraphics* [width=4in,keepaspectratio]{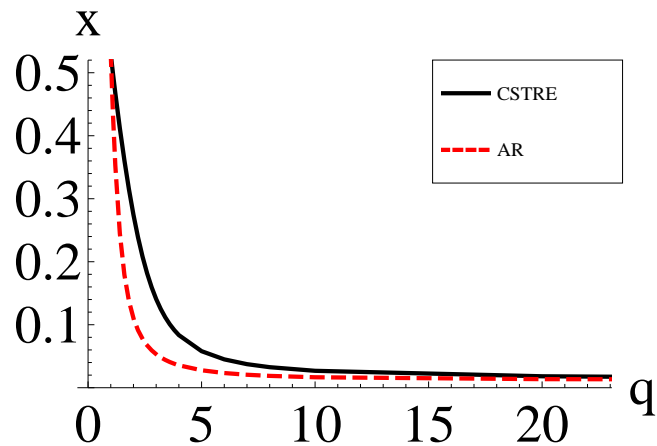}} 
\caption{The comparison between implicit plots of $\tilde{D}^{T}_q(\rho^{(3)}_5(x)\vert\vert \rho_{B}) = 0$ and $ S^{T}_{q}(A\vert B) = 0 $, as a function of $ q $ in the $1:4$ partition of the $5$-qutrit ($N=5$, $d=3$) state $\rho^{(3)}_5(x)$.  A rapid decrease in the value of $ x $, in comparison with  $\tilde{D}^{T}_q(\rho^{(3)}_5(x)\vert\vert \rho_{B})$, can be observed in the case of $ S^{T}_{q} (A\vert B) $.}
\label{abe35}
\end{figure}

\begin{figure}[ht]
\centerline{\includegraphics*[width=4in,keepaspectratio]{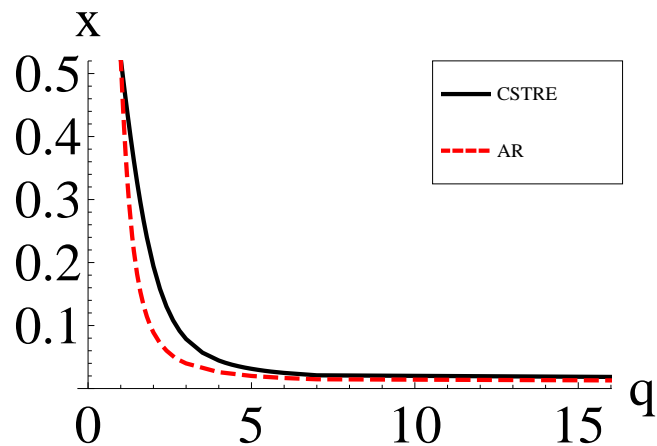}}
\caption{The comparison between implicit plots of $\tilde{D}^{T}_q(\rho^{(5)}_4(x)\vert\vert \rho_{B}) = 0$ and $ S^{T}_{q}(A\vert B) = 0 $, as a function of $q$ for $4$-partite ($N=4$), $5$-level ($d=5$) Werner-Popescu states $\rho^{(5)}_4(x)$.} 
\label{abe54}
\end{figure}

Even though the separability range of $ \rho^{d}_{N}(x)$, obtained using both CSTRE and AR-conditional entropy are same, there is a difference in the way the parameter $x$ converges to the value $x_\infty$, the value of $x$ for which $\mbox{lim}_{q\rightarrow \infty}\,S_q(A\vert B)=0$, $\mbox{lim}_{q\rightarrow \infty}\,\widetilde{D}^{T}_q(\rho^{(d)}_N(x)\vert\vert \rho_{B})=0$.  The rapid convergence of the parameter $ x $ with increasing values of $q$ in the case of AR $q$-conditional entropy is illustrated in Figs. \ref{abe35}, \ref{abe54}.  
\begin{table} 
\caption{The comparison of the value of  $ x $ for $ q = 2 $, obtained through AR-, CSTRE criteria}
\label{ch5table2}
\begin{tabular} {|c|c|c|c|c|c|c|c|c|c|}
\hline
Criterion & \multicolumn{3} { c| } {3-level} & \multicolumn{3} { c| } {4-level} & \multicolumn{3} { c| } {5-level}  \\ 
\cline{2-10}
& 3-party & 4-party & 5-party  & 3-party & 4-party & 5-party  & 3-party & 4-party & 5-party \\
\hline
CSTRE & 0.3837 & 0.3114 & 0.2744 & 0.3108 & 0.2396 & 0.2116 & 0.2610 & 0.1943 & 0.1730  \\
\hline
AR & 0.3162 & 0.1889 & 0.1104 & 0.2425 & 0.1240 & 0.0623 & 0.1961 & 0.0890 & 0.0399   \\
\hline
\end{tabular}
\end{table}  
Table \ref{ch5table2} provides the values of the parameter $x$ at which CSTRE, AR $q$-conditional entropy becomes zero,  when  
$q=2$, for different  $d$ and $N$.
From Table \ref{ch5table2} one can easily note that the parameter $x$ is rapidly decreasing in AR method even for $ q = 2 $ thus confirming its relatively rapid convergence in comparison with that of CSTRE in the limit $q\rightarrow \infty$. 

\begin{figure}[ht]
\centerline{
\includegraphics*[width=4in,keepaspectratio]{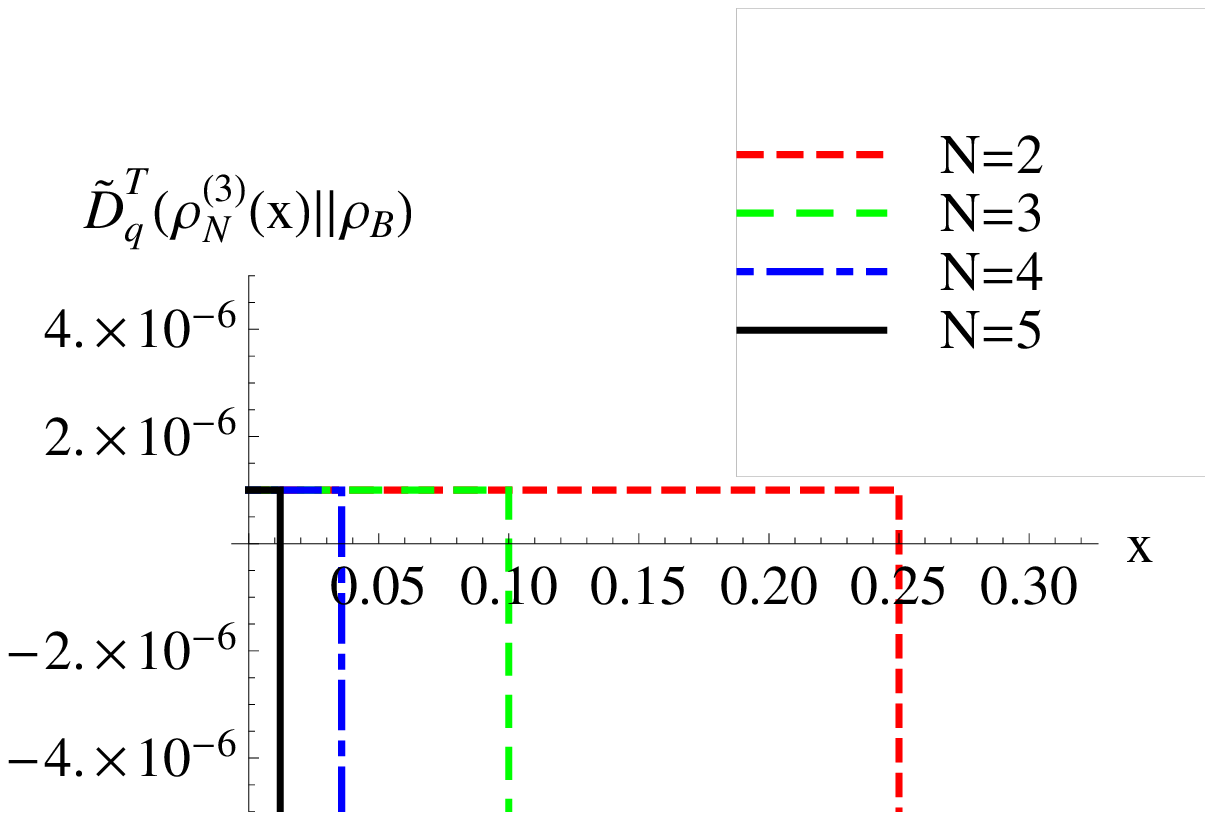}}
\caption{The graph of CSTRE $\tilde{D}^{T}_q(\rho^{(3)}_N(x)\vert\vert \rho_{B})=0$ versus $x$ for different values of $N$ when $d=3$. The decrease of the separability range with $N$, for any given $d$ is clearly seen.} 
\label{d3N}
\end{figure}

\begin{figure}[ht]
\centerline{
\includegraphics*[width=4in,keepaspectratio]{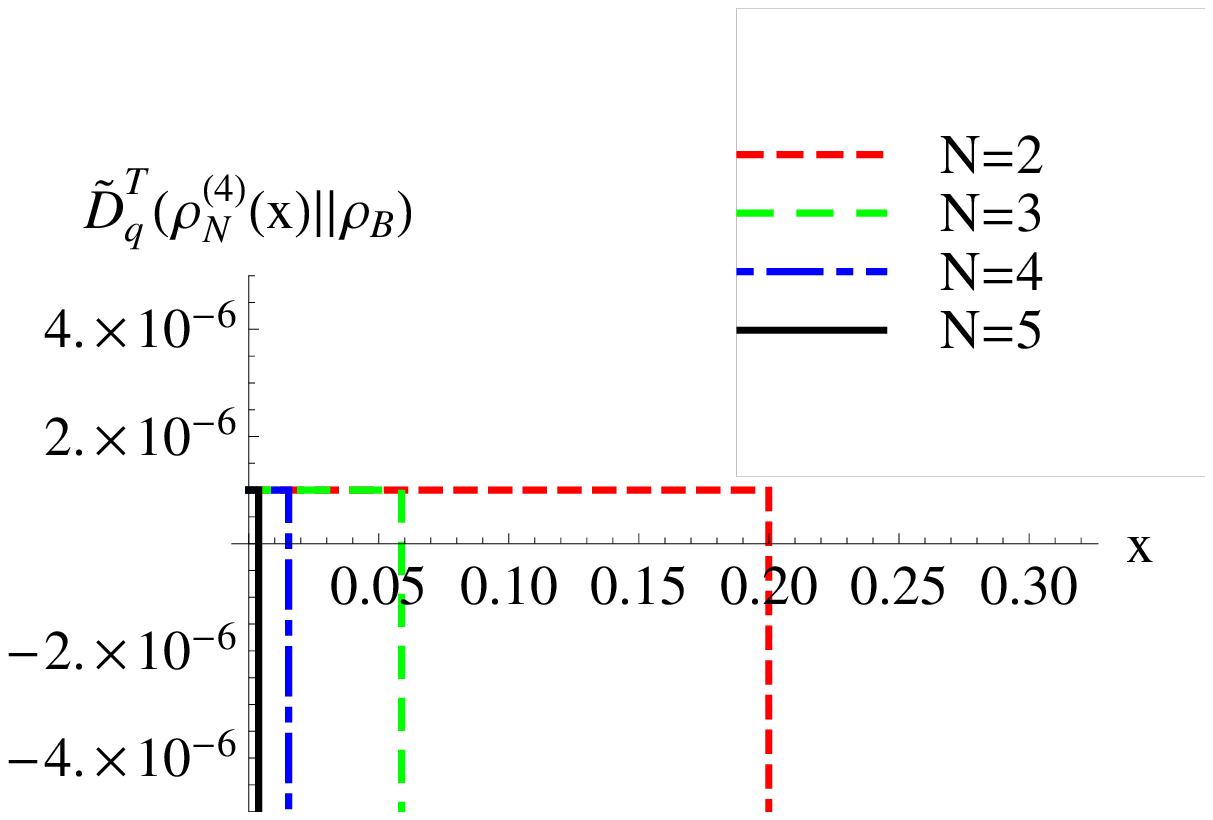}}
\caption{The graph of CSTRE $\tilde{D}^{T}_q(\rho^{(4)}_N(x)\vert\vert \rho_{B})=0$ versus $x$ for different values of $N$ when $d=4$. } 
\label{d4N}
\end{figure}

It is also evident from Table \ref{ch5table2} that the separability range decreases with the number of subsystems i.e., with the increase of $N$ for any given $d$. This feature is illustrated in Figs. \ref{d3N}, \ref{d4N}.
Similarly a comparison of Figs. \ref{d3N}, \ref{d4N} illustrates that for any given $N$, the separability range decreases with increasing $d$. 
Thus a state of the Werner-Popescu family is entangled throughout the parameter range $x$ if its constituents are qudits with larger $d$. More qudits in the state implies a single qudit remains entangled with the remaining $N-1$ qudits in the whole parameter range.

\section{Summary}
\label{ch5:summary}
In this article, the CSTRE criterion is employed to find out the $1:N-1$ separability range of $N$-qudit Werner-Popescu states. It is observed that the $1:N-1$ separability range obtained through both CSTRE and AR $q$-conditional entropy criteria  match with each other for these states. The maximally mixed and hence commuting nature of the single qubit density matrix with the Werner-Popescu state is found to be the reason behind the matching of the $1:N-1$ separability ranges due to commutative AR-criterion and non-commutative CSTRE criterion. The relatively smoother convergence of the parameter $x$ with respect to increasing $q$  is observed in the case of implicit plots of CSTRE in comparison with the convergence in the case of AR $q$-conditional entropy thus establishing the supremacy of CSTRE criterion over the AR-criterion. The $1:N-1$ separability range obtained for $N$-qudit Werner Popescu states using entropic criteria is seen to match with that obtained using an algebraic necessary and sufficient condition for separability.

\section*{Acknowledgements}
Anantha S. Nayak acknowledges the support of Department of Science and Technology (DST), Govt. of India through the 
award of INSPIRE fellowship; A. R. Usha Devi is supported under the University Grants Commission (UGC), India (Grant No. MRP-MAJOR-PHYS-2013-29318



\end{document}